\documentclass{elsart}
\usepackage{graphicx,amssymb}
\newcommand{\av}[1]{\langle{#1}\rangle}
\def\ls{< \kern -12pt \lower 5pt \hbox{$\displaystyle{\sim}$}}
\def\ge{> \kern -12pt \lower 5pt \hbox{$\displaystyle =$}}
\def\le{< \kern -12pt \lower 5pt \hbox{$\displaystyle =$}}
\def\gs{> \kern -12pt \lower 5pt \hbox{$\displaystyle{\sim}$}}

\def\gdot{\dot{\gamma}}
\def\Tve{{\stackrel{\leftrightarrow}{\sigma}_{\rm p}}}
\def\TW{{\stackrel{\leftrightarrow}{W}}}
\def\asigma{{\stackrel{\leftrightarrow}{\sigma}}}
\def\EQ{\begin{eqnarray}}
\def\EN{\end{eqnarray}}
\def\bv{{\bf v}}
\def\rs{{\dot\gamma}\tau}
\def\bk{{\bf k}}
\def\br{{\bf r}}

\def\nb{{\nabla}}
\def\dfrac#1#2{{\displaystyle\frac{#1}{#2}}}
 
\def\be{\begin{equation}}
\def\en{\end{equation}}
\newcommand{\bi}[1]{\mbox{\boldmath$#1$}}
\def\p{\partial}
\def\bea{\begin{eqnarray}}
\def\ena{\end{eqnarray}}

\begin{document}
\begin{frontmatter}
\title{Spatio-temporal structures in sheared polymer systems}
\author{Akira Furukawa and Akira Onuki}
\address{Department of Physics, Kyoto University, Kyoto 606-8502, Japan}

\begin{abstract}
We investigate spatio-temporal structures in sheared polymer systems 
by solving a time-dependent Ginzburg-Landau model in two dimensions. 
(i) In polymer solutions above the coexistence curve, 
crossover from  linear to nonlinear regimes 
occurs  with increasing the shear rate.  In the nonlinear regime 
the solution behaves chaotically with large-amplitude 
composition fluctuations. A 
characteristic heterogeneity length is calculated 
in the nonlinear regime.     
(ii) We also study dynamics of shear-band structures  
 in  wormlike micellar solutions under the condition of fixed stress. 
The average shear rate exhibits large temporal fluctuations 
with occurrence of large disturbances in the  spatial structures.  
\end{abstract}
\begin{keyword}
Two-fluid model, 
Shear-induced concentration fluctuations, 
Stress-diffusion coupling, 
Shear-banding, 
Rheological chaos.
\end{keyword}

\end{frontmatter}

\section{Introduction}
In complex fluids such as 
polymer solutions, liquid crystals, or  
surfactant systems, 
the rheological properties can be very singular and complex. 
Application of flow field to  complex fluids 
often  gives rise to dramatic structural changes, 
sometimes coupled to phase separation  \cite{Larson,Onukibook}. 
Even chaotic spatio-temporal structures 
have  been observed  
far away from equilibrium.

In this paper  
we will present  numerical results on 
two sheared polymer systems:     
(i) In polymer solutions above the coexistence curve,  
 application of flow has  been observed 
to induce strong composition heterogeneities 
\cite{Wu,Hashi1,Han,Saito,Egmond,Moses,Boue,Pine2}. 
The origin is now ascribed to a dynamical coupling 
between stress and diffusion in asymmetric 
viscoelastic mixtures 
\cite{Helfand,OnukiPRL,Milner,Doi-Onuki}.
This is in marked contrast to  
shear-induced homogenization or mixing 
observed in near-critical fluids \cite{Onuki1,Beysens} and 
in polymer mixtures with small entanglement 
effects \cite{Hashimoto_ternary,StringPRL}. 
Because some  numerical results have already 
been presented 
\cite{Onuki97,PRE2004,Okuzono,Yuan-Kawakatsu}, 
we will here focus our attention  on  crossover 
from  linear to nonlinear regimes with increasing the shear rate 
and chaotic behavior in the nonlinear regime.    
Furthermore, viscoelastic phase separation below 
the coexistence curve 
in shear flow has recently been 
studied, where polymer-rich domains are fragmented 
like broken gels \cite{PRE2004}. 
(ii) Next we will briefly examine 
shear  effects in  wormlike micellar solutions 
where the stress-strain relation is nonmonotonic. 
Wormlike micells have significant chain 
flexibility and can break and combine \cite{Cates}. 
In such systems 
shear bands composed of low-shear regions 
and high-shear regions parallel to flow have
 been observed 
\cite{SB-experiment1,Pine,SB-experiment2,SB-experiment3} 
and  {\it rheological chaos}, 
which occurs  at low Reynolds numbers, 
has been discussed experimentally 
\cite{RC-experiment1,RC-experiment2,RC-experiment3,RC-experiment4,RC-experiment5} 
and theoretically 
 \cite{RC-theory0d-1,RC-theory1d-1,a,b}.   
In the presence of shear bands 
 macroscopic observables, 
such as the shear stress at fixed  shear rate 
or the shear rate at fixed  shear stress,  
exhibit  irregular 
 nonstationary  {\it chaotic}  behavior,  
which arises from large-scale disturbances of the shear bands.     
However, not enough  understanding 
has  yet been attained on
 such  nonstationary behavior of the shear bands. 
In this paper we use the word {\it chaotic} loosely 
without rigorous characterization.

Our numerical simulations will be performed in a Ginzburg-Landau scheme, 
where a conformation tensor representing the chain deformations 
is a dynamic variable. In previous papers 
numerical simulations  (still in two dimensions)  
in this line have provided useful 
insights into the complex nonlinear shear effects 
\cite{Onuki97,PRE2004,Okuzono,Yuan-Kawakatsu}.
The organization of this paper is as follows. 
We will explain our theoretical scheme in Section 2.  
We will then present  numerical results  for polymer solutions 
in Section 3 and for  wormlike micellar solutions in 
Section 4.

\section{Model equations in dimensionless forms}

Phase behavior of polymer solutions 
is usually  described in terms of the polymer 
volume fraction $\phi$ only \cite{PGbook}. 
However, to  describe  the viscoelastic effects, 
it is convenient to introduce a symmetric 
tensor  ${\TW}=\{ W_{ij}\}$ which  represents 
chain conformations 
undergoing deformations. 
The free energy density $f$  then depends on 
$\phi$ and $\TW$. For simplicity, we will  measure 
the free energy density and the stress  
in units of $\sigma_0=   k_{\rm B} T/{v_0}N^{3/2}$ 
where $v_0$ is the monomer volume 
and  $N$ is the monomer number on a chain. 
In terms of the scaled polymer density 
$\Phi=N^{1/2}\phi$,  $f$  reads 
\be 
f = \Phi \ln \Phi -\frac{u}{2}\Phi^2 
+\frac{1}{6}\Phi^3 
 + \frac{2}{\Phi}
|\nabla\Phi|^2 + \frac{g}{4}  \Phi^3 Q (\TW), 
\en 
in the semidilute region $\phi \ll 1$ 
 close  to the 
coexistence curve (in theta solvent). 
The  space  
integral $F= \int d{\bi r} f$ gives the free energy.  
The first three terms constitute the Flory-Huggins 
free energy  and 
the fourth term represents 
the gradient free energy, where  
$u$  is  related to the interaction parameter $\chi$ 
by $u=N^{1/2}(2\chi-1)$ \cite{PGbook}. We have $u=2$ and 
$\Phi=1$  at the critical point and $u=\Phi+\Phi^{-1}$ 
on the spinodal curve. 
We  measure space  in units of 
$\ell=  {bN^{1/2}}/{2 \sqrt{18}}$ ($\sim$ 
gyration radius),  where 
$b=v_0^{1/d}$ is the monomer length with $d$ being  
the space dimensionality. 
The equilibrium average of 
$W_{ij}$ is the unit 
tensor $\delta_{ij}$ and the   deviation 
$\delta W_{ij}= W_{ij}- \delta_{ij}$ 
gives rises to an increase in the free energy.
We adopt  the Gaussian form 
$Q(\TW)= \sum_{ij} (\delta W_{ij})^2$. 
The coefficient $g$ is of order 1 in theta solvent 
and will be set equal to 1.

Next we  derive 
the dynamic  equations using a two-fluid model.  
We assume that  the polymer and the solvent have 
different velocities, 
${\bv}_p$ and ${\bv}_s$, 
respectively.  
The continuity equation is written as 
${{\partial}\phi/{ \partial t}} =-
\nabla\cdot (\phi {\bi v}_{\rm p} )$. 
In the semidilute region it is rewritten  as 
\be 
{\frac{\partial}{ \partial t}}\Phi 
+\nabla \cdot (\Phi{\bi v}) =
-\nabla\cdot(\Phi{\bi w}),
\label{eq:2.14}
\en 
in terms of the  volume-averaged velocity
 $\bv=\phi\bv_p+(1-\phi)\bv_s$   and 
the relative velocity ${\bi w}=\bv_p-\bv_s$. 
For $\phi \ll 1$ we have $\bv_s \cong \bv$ 
and $\bv_p \cong \bv+{\bi w}$.
The right hand side of Eq.(2) represents the diffusion.  
We assume that the solution is 
incompressible or $\nabla\cdot{\bi v}=0$. 
The motion of the conformation tensor $\TW$ is driven by 
the polymer velocity ${\bi v}_p$ as 
\cite{Larson,Yuan-Kawakatsu,Jhonson-Segalman,Yuan,Ball,Olmsted-Fielding}  
\EQ
&&(\dfrac{\partial}{\partial t}+{\bv}_p\cdot\nabla)W_{ij}
-\sum_{k}\biggl[a\bigl(S_{ik}\delta W_{kj}+\delta W_{ik}S_{kj}\bigr)
+\bigl(\delta W_{ik}\Omega_{kj}-\Omega_{ik}\delta W_{kj}\bigr)\biggr]\nonumber\\&&=
(\nb_j v_{{\rm p}i}+\nb_i v_{{\rm p}j})
-({1}/{\tau}-{\mathcal D}\nb^2) \delta W_{ij}, 
\EN
where $\nabla_j =\p/\p x_j$, 
$S_{ij}=(\nb_i v_{{\rm p}j}+\nb_j v_{{\rm p}i})/2$, 
$\Omega_{ij}=(\nb_i v_{{\rm p}j}-\nb_j v_{{\rm p}i})/2$, 
and $\tau$ is the very long rheological time. 
The left hand side is the frame-invariant 
time derivative of the tensor 
 $\delta W_{ij}$ and $|a|\le 1$ will be assumed 
 in Section 4.  The 
${\mathcal D}$ is  a kinetic coefficient  
 originally introduced  to explain 
 shear banding in the 
Johnson-Segalman model 
\cite{Yuan,Ball}, 
but is not indispensable 
in the presence of the gradient  term 
in the free energy Eq.(1) for the composition. 
In our previous papers 
we have assumed $a=1$ and ${\mathcal D}=0$ for polymer solutions.
However, we will assume $a=0.4$ and ${\mathcal D}=1$ for 
wormlike micellar solutions. 
The network (viscoelastic) stress due to $\delta W_{ij}$ 
consistent with Eq.(3) is given by \cite{Onukibook}  
\EQ
\sigma_{{\rm p}ij}=G\sum_{k}\delta W_{ik}(\delta_{kj}+a\delta W_{kj}) 
+\frac{1}{4}GQ\delta_{ij},
\label{viscoelastic-stress} 
\EN
where $G=g\Phi^3$ has the meaning of the 
shear modulus (in units of $\sigma_0$)
 for small rapid deformations. 
The linear viscosity is given by $\eta=\eta_0+ 
G\tau$ in  homogeneous states under weak steady shear, which  
is taken to be much larger than the solvent viscosity 
$\eta_0$.  Note that  if   $G\gg 1$, 
large viscoelastic effects 
are produced even for small  $\delta W_{ij}$. 
The eigenvalues of 
$\delta W_{ij}$ remain small (at most of order 0.1)   
in our simulations \cite{PRE2004},  justifying  the 
Gaussian form of $Q$.  
Physically, this means that 
 the chains are  weakly stretched in shear flow. 
 In our case  
we simply have $\sigma_{{\rm p}ij}\cong 
G\delta W_{ij}$ from  Eq.(4).

Hereafter time will be measured in units of 
${\tau_0}= {\eta_0}/{4\sigma_{0} } (\sim 6\pi\eta_0
 \ell^3/k_{\rm B}T$). 
The velocities will be   measured in units of 
$\ell/\tau_0=4\sigma_0\ell/\eta_0$. If we neglect the inertial 
effects (accelerations) in the two-fluid model, 
the velocities $\bi w$ and $\bi v$ read  \cite{PRE2004} 
\be
{\bi w}= 
-\nabla\frac{\delta F}{\delta \Phi}+  
\frac{1}{\Phi}\nabla\cdot\Tve
-\frac{3}{4} g{Q}{\Phi}
\nb \Phi , \qquad   
{\bi v}=  
 \av{\bi v}- \frac{1}{4\nabla^2} 
[\Phi {\bi w}  ]_\perp ,
\label{eq:2.48}
\en 
in the semidilute 
region. The $\av{\bi v}$ is the mean  shear flow, 
$[\cdots]_\perp$ denotes taking the transverse part 
(whose Fourier component is perpendicular to the 
wave vector)  and the inverse operation 
$1/{\nabla^2}$ may 
be expressed in terms of the  Oseen tensor 
in the limit of large system size \cite{Onukibook}. 
 The stress force 
$\nabla\cdot\Tve$ in the diffusion flux $\Phi{\bi w}$ 
gives rise to the so-called stress-diffusion coupling 
\cite{Onukibook,Doi-Onuki}. 
This means that diffusion is induced by 
stress imbalance 
$\nabla\cdot\Tve \neq {\bi 0}$, which leads to   
various viscoelastic effects 
such as nonexponential decay in dynamic light scattering 
or shear-induced phase separation.

We now have a closed set of dynamic equations 
for  $\Phi$ and $\TW$. 
In particular,   use of these equations  yields  
the time derivative of the free energy  $F$, 
\be
\frac{d F}{dt} =
\int d{\bi r} \bigg [ \frac{\delta F}{\delta \Phi}
\frac{\p \Phi}{\p t} + \frac{G}{2} 
\sum _{ij} \delta W_{ij}
\frac{\p W_{ij}}{\p t} \bigg ] 
= \int d{\bi r} \bigg [  -\Phi^2{\bi v}_{\rm p} \cdot {\bi w}    
  - \frac{G}{2\tau}Q \bigg ]. 
\label{eq:2.29}
\en
If $\av{\bi v}={\bi 0}$, we have  $dF/dt \le 0$. 
The steady  condition $dF/dt=0$ is 
realized  in equilibrium where 
$\delta F/\delta \Phi={\rm const}.$ 
and $\delta W_{ij}= 0$.  This  is 
 a self-consistent condition of 
dynamic equations which have stable equilibrium 
solutions \cite{Onukibook}.   Under this condition 
we may add the Langevin thermal noise terms 
in the dynamic equations Eqs.(2) and (3), 
although they have been  omitted for simplicity.

Assuming that the flow direction is along 
the $x$ axis and the velocity-gradient direction is 
along the $y$ axis,  we give the expressions for 
 the shear stress  $\sigma_{xy}$ and the normal stress 
difference $N_1= \sigma_{xx}- \sigma_{yy}$  
in units of $\sigma_0$ \cite{PRE2004}, 
\bea 
\sigma_{xy}&=& \sigma_{{\rm p}xy}- {4}{\Phi}^{-1}\nabla_x\Phi\nabla_y\Phi 
 +4 (\nabla_x v_y+ \nabla_y v_x) ,
\\ 
N_1 &=& N_{{\rm p} 1}- {4}{\Phi}^{-1} 
(\nabla_x\Phi\nabla_x\Phi - \nabla_y\Phi\nabla_y\Phi)
+8 (\nabla_x v_x- \nabla_y v_y) ,
\label{eq:2.58}
\ena 
with $\bi v$ in units of $\ell/\tau_0$.  
The first terms  are the network  stress components, 
the second ones arises from 
the gradient free energy (yielding  the surface tension 
 contributions in two-phase states), and the 
third ones are the viscous contributions. 
In homogeneous states with 
 weak shear $\gdot= \av{\nabla_y v_x}$ 
we find $\sigma_{xy}= (g\Phi^3\tau + 4)\gdot$ with 
$\tau$ in units of $\tau_0$, so 
$\eta/\eta_0=1+g\Phi^3\tau/4 \gg 1$.  
For our inhomogeneous flows below, we also 
have $\av{\sigma_{ij}}\cong \av{\sigma_{{\rm p}ij}}$ 
and $\av{N_{1}} \cong \av{N_{{\rm p}1}}$.

\section{Nonlinear structures  in semidilute 
polymer solutions}

\subsection{Numerical method for sheared fluids}

For polymer solutions we set $a=1$,  ${\mathcal D}=0$, and 
 $\tau=\tau_0(0.2+\Phi^4)/(1+Q)$, 
where $Q$ in the denominator gives rise to 
shear-thinning for $\gdot\tau \gs 1$ 
even in homogeneous states 
\cite{Onukibook,PRE2004}.  For weak chain-stretching in the semidilute region, 
we have $\tau/\tau_0 \cong \Phi^4 \gg 1$.   
Here we explain our 
numerical method. 
We integrate the dynamic equations in two dimensions on a 
$256\times 256$ lattice. Our 
numerical scheme  is to use  
 the deformed coordinates,
\be 
x'=x-\gamma(t) y, \qquad  y'=y, 
\en 
 under  the periodic boundary condition 
$f(x',y'+L)=f(x'+L,y')=f(x',y')$ for  any quantity $f(x',y')$ 
in the deformed-space representation  
\cite{Onuki_comp}.  Then we can use the FFT method in 
the deformed space to 
calculate $\bi v$ from Eq.(5).  
Here we take  $\gamma (t)=\gdot t$ 
at  constant  $\gdot$.  In Section 4, however,  $\gamma(t)$ 
 will  be determined such that the 
average shear stress $\av{\sigma_{xy}}$ is held at 
a given constant. 
As the initial condition at $t=0$ we assign Gaussian 
random compositions at each lattice point, with  mean
value $\av{\Phi}$ and variance 0.1. 
For $t>0$ we  solve the dynamic equations  
without the Langevin  noise terms.

\subsection{Mechanism of shear-induced fluctuation enhancement}

For very small  shear rate $\gdot \ll \tau^{-1}$ 
the  linear stability analysis  predicts 
the composition fluctuation enhancement 
above the coexistence curve owing to 
the stress-diffusion coupling 
\cite{Helfand,OnukiPRL,Milner,Doi-Onuki}.  
At very long wavelengths  the 
relaxation rate  strongly 
depends on the direction 
of the wave vector $\hat{\bi k}= k^{-1}{\bi k}$ as   \cite{Onukibook}
\be 
\Gamma({\bi k})= \frac{Dk^2}{1+k^2\xi_{\rm ve}^2} 
\bigg [1 -  \frac{2\phi}{K_{\rm os}}\frac{\p \eta}{\p \phi}
 \gdot \hat{k}_x \hat{k}_y \bigg ],
\en  
where $D$ is the mutual diffusion constant, 
$K_{\rm os}=\sigma_0(\Phi-u\Phi^2+\Phi^3) $ is the osmotic modulus, 
and $\xi_{ve} (\sim ({\eta/\eta_0})^{1/2}b/\phi)$ is 
the viscoelastic length much longer than the correlation length 
($\sim b/\phi$). The reduction of the relaxation rate (growth 
rate in spinodal decomposition)  or the Onsager kinetic coefficient 
 by the factor 
$1/(1+k^2\xi_{\rm ve}^2)$  has been detected by  scattering 
 experiments without shear \cite{Sc,Toyoda}. 
Note that $\Gamma({\bi k})<0$ 
in   the direction of $\hat{k}_x = \hat{k}_y$ 
for $K<\gdot \phi \p\eta/\phi$. In  the semidilute region 
 $\p \ln \eta/\p \ln \phi 
\sim 6$ experimentally 
(7 in our simulation since $\eta\cong  G\tau$). 
It then  follows 
the condition of strong 
fluctuation enhancement,  
\be 
\gdot > K_{\rm os}/ 6\eta \sim 0.1\tau^{-1}, 
\en   
in fair agreement with a scattering experiment 
\cite{Saito}. Under Eq.(10) the linear theory cannot 
be used and we encounter  a nonlinear  regime. 
The orign of the fluctuation enhancement is as follows: 
when  the viscoelastic 
properties strongly depend on $\phi$, appearance 
of inhomogeneities  
can lower  the network  free energy ($\propto g$ 
in Eq.(1))  and the stress (at fixed  shear rate).  
By the same reason, when long and short polymers pass 
through a capillary tube, the longer ones 
migrate  toward  the tube 
center \cite{Doi-Onuki}.

\subsection{Shear-induced fluctuations above the coexistence curve}

Figure 1 shows time evolution of the  variance 
$\sqrt{{\langle(\delta\Phi)^2\rangle}(t)}$, 
the shear stress $\av{\sigma_{xy}}(t)$, 
and the  normal stress difference $\av{ N_1} (t)$ 
after application of shear at $t=0$ for 
$\dot\gamma=0.05 \cong 0.8/\tau$. The averages are taken over 
all the lattice points. Here the system is 
above the coexistence curve  with $u=2$ 
 and $\langle\Phi\rangle=2$. 
Though there is no thermal noise in the dynamic equations, 
the variance saturates at about $0.6$ in the late stage. 
Figure 2 suggests that 
the  system is in a state of deterministic chaos,  
undergoing partial  or  incomplete phase separation.  
Both  $\av{\sigma_{xy}}(t)$ and $\av{N_1}(t) $ 
exhibit a pronounced peak at the onset of fluctuation enhancement 
at $t \sim 25$ and noisy behavior 
follows in the subsequent dynamical steady state. 
The noise of these quantities arises from breakup of the polymer-rich  
domains supporting the stress. 
The effective viscosity $\eta_{\rm eff}= \langle{\sigma_{xy}}\rangle
/\dot\gamma$ in the steady state is about $8$ times larger 
than the solvent viscosity $\eta_0$. 
It would be $28\eta_0$ for the homogeneous case, 
so marked shear thinning is realized.  
For comparison the shear stress (upper broken line) 
and the normal stress difference (lower broken line) 
are calculated for the homogeneous case.
\begin{figure}[bht]
  \begin{center}
    \includegraphics[width=80mm]{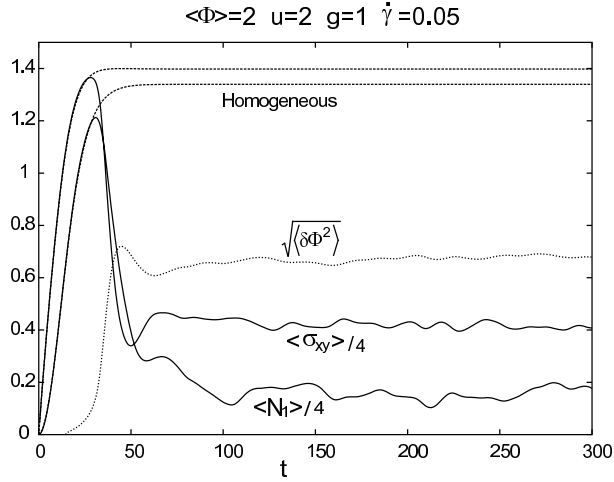}
  \end{center}
  \caption{
    Time evolution of  $\sqrt{\langle\delta\Phi^2\rangle}$ (dotted line), 
    $\langle\sigma_{xy}\rangle$, and $\langle N_1\rangle$ (solid lines) 
    after application of shear.
    At $t=0$, the system is above the coexistence curve. 
    The shear-induced fluctuations (shown in Fig.2)  
    give rise to the pronounced stress overshoots. 
    The upper curves are those for the homogeneous case. 
    }
\end{figure} 

\begin{figure}[b]
  \begin{center}
    \includegraphics[width=90mm]{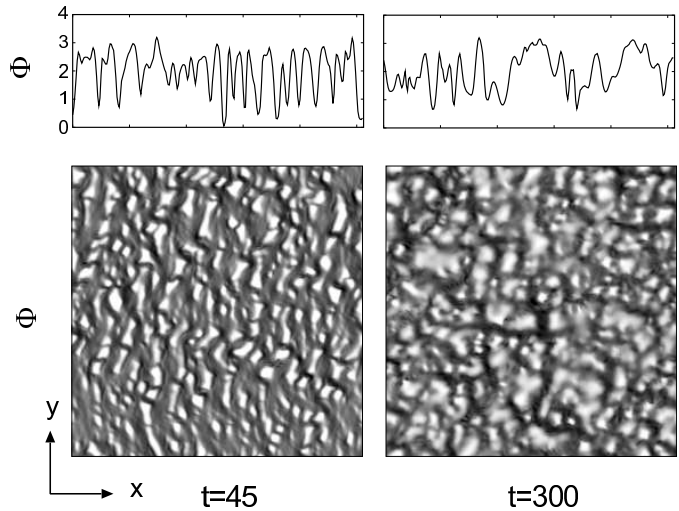}
  \end{center}
  \caption{
    $\Phi(x,y,t)$ 
    at $t=45$ (left) and $t=300$ (right)  above the coexistence curve. 
    The top figures are the profiles along   the  $x$ axis 
    at $y=128$. The parameters are the same as those 
    in Fig.1. 
    For  $t>100$ a dynamical steady state is realized.}
\end{figure} 
Figure 2 displays the normalized composition  
$\Phi({\bi r},t)$ 
at $t=45$ (left) and $t=300$ (right) 
in the run which produced Fig.1. 
In the top figures we cannot  see well-defined 
interfaces, but the composition changes over a 
mesoscopic, characteristic  length $R_{\rm h}$ 
 in the steady state.  
We numerically find that the average gradient term 
$\av{2|\nabla\Phi|^2/\Phi}$ 
and the average elastic energy $\av{g\Phi^3 Q/4}$ 
in the dimensionless free energy density Eq.(1)   
are of the same order for various $\gdot$ and $\av{\Phi}$. 
We estimate the former as $\av{\Phi}/R_{\rm h}^2$ 
and the latter as $\av{\sigma_{xy}}^2/\av{\Phi}^3$ to obtain 
\EQ
R_{\rm h} \sim \av{\Phi}^2/\av{\sigma_{xy}},  
\label{characteristic_size}
\EN 
or $R_{\rm h} \sim 
b G/\phi\av{\sigma_{xy}}$ in the original units.  
The behavior of $\av{\sigma_{xy}}$ 
as a function of $\gdot$ can be known from Fig.5. 
On the other hand, to the stress, 
the  nework contributions are mostly much larger 
than the  gradent contributions in  Eqs.(7) and (8).

The length $R_{\rm h}$ should be measurable by scattering.  
In Fig.3 the time-averaged structure factor 
$S(k_x,k_y)=\av{|\Phi_{\bk}|^2}$ in 
 the dynamical steady state is shown.
It   has sharp  peaks in the $k_x$-$k_y$ plane 
close to the $k_x$ axis, presumably at $k_x \sim \pm 2\pi/R_{\rm h}$. 
This is  in agreement with 
 the experimental structure factor for $\gdot \tau \sim 1$ 
in the  $k_x$-$k_y$ plane \cite{Wu}. 
It is worth noting that the light scattering  in the 
  $k_x$-$k_z$ plane was also intensified along the 
  $k_x$-axis  for $\gdot \tau \sim 1$ 
\cite{Saito}. Microscope pictures 
in the $x$-$z$ plane of the shear-induced composition 
fluctuations closely resemble the pattern  at $t=300$ 
in Fig.2 \cite{Moses}.  
These suggest  stronger correlations 
 perpendicular to the 
flow direction  than along  
 the flow  for $\gdot \tau \sim 1$.  
Below the coexistence curve also,  
the  structure factor exhibits similar 
behavior \cite{PRE2004}.  
\begin{figure}[htb]
  \begin{center}
    \includegraphics[width=90mm]{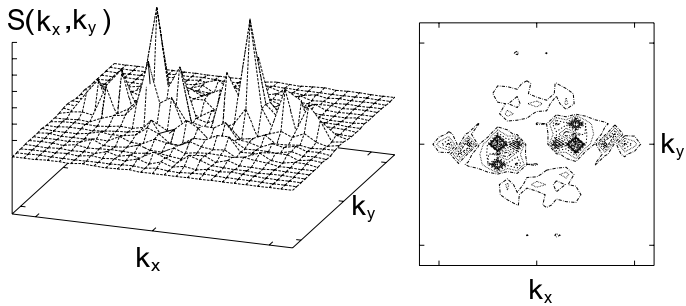}
  \end{center}
  \caption{ 
    Bird-view (left) and contour plot (right) of 
  the time-averaged structure factor 
    $S(k_x,k_y)=\langle|\Phi_{\bk}|^2\rangle$ 
   in the dynamical steady state in Fig.2. 
   }
\end{figure}

\subsection{Chaotic behavior above the coexistence curve}


As shown in Fig.4, there emerge two dynamic 
regimes at longer times. 
When $\gdot$ is smaller than a characteristic 
shear rate $\gdot_{\rm nl} (\cong 0.01$ for $\av{\Phi}=2$), 
the fluctuations enhanced in the early stage   
slowly relax to zero 
due to the convective  deformation by  shear 
in the absence of the thermal 
noise. In this {\it linear} or {\it thermal} regime,  
the steady-state fluctuations can be attained only 
in the presence of the thermal noise. 
It should be noted that  
in the previous theories  
the fluctuation intensity 
was calculated in the linear approximation 
in the presence of the thermal noise. 
\cite{Helfand,OnukiPRL,Milner,Doi-Onuki}.  
In the  {\it nonlinear} or {\it chaotic}  
regime $\gdot>\gdot_{\rm nl}$, 
the large-amplitude fluctuations are continuously produced 
by the nonlinear interactions among 
them even without the thermal noise. 
In Fig.5 we show a discontinuous change 
in  the steady-state variance 
with varying $\gdot$ for three $\av{\Phi}$ (left), 
 which occurs roughly at 
 $\gdot\sim0.2/\tau$ with 
$\tau\cong \tau_0(0.2+\Phi^4)$. 
The variance goes to zero in the linear regime in  
the absence of the  thermal noise. 
Of course, the changeover becomes continuous 
in the presence of the thermal noise. 
In Fig.5 we also show  the normalized effective viscosity 
$\eta_{\rm eff}/\eta_0 
= \av{\sigma_{xy}}/\eta_0\gdot$  versus  
     $\gdot$ (right), which exhibits Newtonian behavior 
in the linear regime and strong shear-thinning behavior 
with $\eta_{\rm eff} 
\sim 1/\gdot$ in the nonlinear regime.  
\begin{figure}[h]
  \begin{center}
    \includegraphics[width=85mm]{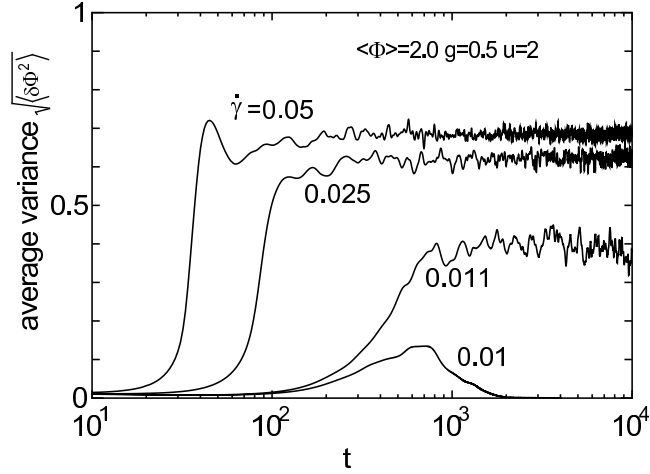}
  \end{center}
  \caption{
    Time evolution of 
    $\sqrt{\av{(\delta\Phi)^2}}(t)$  
    for four values of  $\gdot$  without the  Langevin noise. 
    A chaotic  steady state is realized for the upper 
    three curves. For the smallest shear $\gdot=0.01$ 
    the fluctuations enhanced in the early stage   
    slowly relax to zero. 
  }
\end{figure} 
\begin{figure}[h]
  \begin{center}
    \includegraphics[width=120mm]{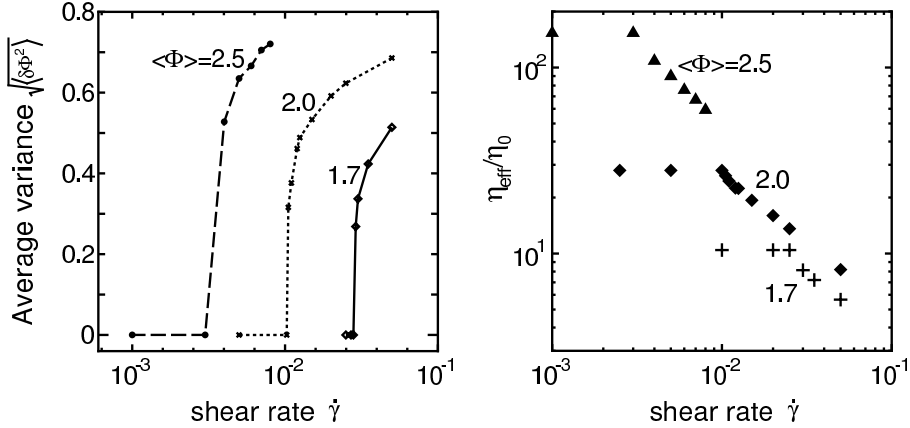}
  \end{center}
  \caption{ Steady-state variance 
    $\sqrt{\av{(\delta\Phi)^2}}$ (left) 
and normalized effective viscosity $\eta_{\rm eff}/\eta_0 
$ (right) versus  
     $\gdot$  
    for three  values of  $\langle\Phi\rangle$. 
Here there is no Langevin  noise terms 
and a sharp discontinuous transition occurs between 
the linear and nonlinear regimes. Shear-thinning 
behavior is marked in the nonlinear regime. 
  }
\end{figure}

We give some remarks. 
(i) In  one-phase states of 
 near-critical fluids without viscoelasticity, 
the critical fluctuations are always suppressed by shear flow 
below the equilibrium level and the thermal noise is   
indispensable to produce the steady-state structure factor  
\cite{Onukibook,Onuki1}. 
(ii) Below the coexistence curve, on the contrary, 
 the random noise terms are generally irrelevant in producing 
large-scale phase-separation patterns and are 
usually omitted in simulations without and with shear flow 
\cite{Onukibook}.  
However small the shear rate is,  
the domains  are eventually 
deformed nonlinearly by  applied shear flow 
as the domains grow. Therefore, 
the linear shear  regime is nonexistent 
in late-stage phase separation. 
For Newtonian  fluid mixtures  with small surface tension,  
  string-like domain structures can be formed  
in steady states \cite{StringPRL}.     
(iii)The system considered here has  very low Reynolds numbers, 
so that the fluid inertia plays no role 
in the irregular patterns.

\section{Nonlinear structures in wormlike micellar solutions}
\subsection{Nonmonotonic stress-strain relation}

\begin{figure}[htb]
  \begin{center}
    \includegraphics[width=80mm]{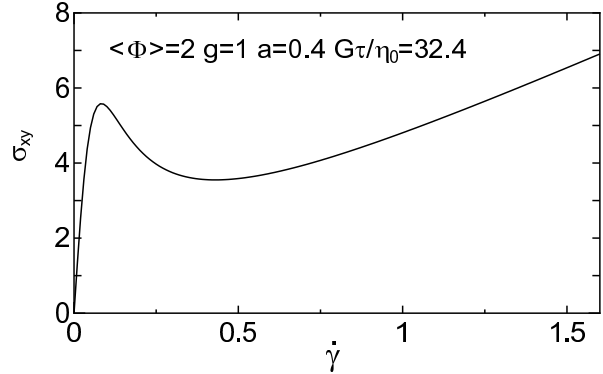}
    \label{x}
  \end{center}
  \caption{ Stress-strain relation Eq.(12) 
for wormlike micellar solutions in homogeneous states.  
 The region $\p \sigma_{xy}/\p \gdot<0$ is  unstable and 
shear bands appear. }
\end{figure}

We present preliminary results on sheared wormlike micellar solutions 
where $a=0.4$, ${\mathcal D}=1$, and $\tau=\tau_0(0.2+\Phi^4)$ in Eq.(3).    
Several authors have used the framework of the two-fluid dynamics 
including the composition  
\cite{Yuan-Kawakatsu,Olmsted-Fielding}.  
For $|a|<1$  it is important that the  
stress-strain curve is nonmonotonic. That is, for 
homogeneous shear flow, Eq.(3) is solved to give 
 $\delta W_{xy}= \rs/[1+(\rs)^2(1-a^2)]$ and  
$\delta W_{xx}/(a+1) = \delta W_{yy}/(a-1)= 
\rs \delta W_{xy}$.  Then Eq.(4) gives  
\EQ
\sigma_{xy}=\eta_0{\dot\gamma}
+G\rs\dfrac{\bigl[1+(\rs)^2(1+a^2)\bigr]}{\bigl[1+(\rs)^2(1-a^2)\bigr]^2}.  
\label{flow_curve}
\EN
For $\eta_0/G\tau \ll 1$ 
the curve has a maximum at $\gdot \sim 1/\tau$ 
and a minimum at $\gdot \sim (G/\eta_0\tau)^{1/2}$ for 
$a$ not very close to 1. 
Fig.6 displays the curve for 
$G\tau/\eta_0=32.4$. 
It is known that  homogeneous states are 
unstable for  
$\p \sigma_{xy}/\p \gdot<0$. Shear banding instability 
occurs in the region of $\sigma_{xy}$ where 
$\gdot$ is multi-valued.

\subsection{Numerical method  at fixed stress}

For this system there have been experiments both at constant shear rate 
\cite{RC-experiment1,RC-experiment2,RC-experiment3,RC-experiment4} 
and at constant shear stress \cite{RC-experiment5}. 
Since simulations of a model similar to ours have already 
been performed at constant shear rate for $a=0$ \cite{Yuan-Kawakatsu}, 
we here show  results at constant shear stress. 
We solve the dynamic equations in 
Section 2 on a $128\times 128$ lattice using the 
numerical scheme in Section 3 \cite{Onuki_comp}.  
We impose the periodic boundary condition in 
the deformed space, so the 
boundary effects do not appear in our scheme. 
We use  the Stokes approximation 
$\nabla\cdot {\asigma}={\bi 0}$  
to find 
$\int_0^{L} dx \nabla_y \sigma_{xy}({\bi r},t)=0$, 
where $\asigma$ is the (total) stress tensor 
and  $L$ is the system length in the $x$ axis.  
Then the lateral average defined by  
\be 
\Sigma=  \int_0^{L} dx  \frac{1}{L} 
\sigma_{xy}({\bi r},t)
\en 
becomes independent of $y$. If $\Sigma$ is 
a constant independent of time, 
Eq.(7) yields the  average shear rate  dependent on  time  
in the dimensionless form,  
\EQ
4\av{\dot\gamma}(t)=\Sigma -\av{\sigma_{{\rm p}xy} -4\Phi^{-1}
\nb_x\Phi\nb_y\Phi},
\EN   
where $\av{~~}$ denotes taking the space average.  
At each time step we calculate the average strain 
 $ \int_0^t  dt' \av{\gdot}(t')$ 
as the shear  strain $\gamma(t)$ in Eq.(9).

\subsection{Chaotic behavior  
with concentration inhomogeneities at fixed stress}

\begin{figure}[h]
  \begin{center}
    \includegraphics[width=95mm]{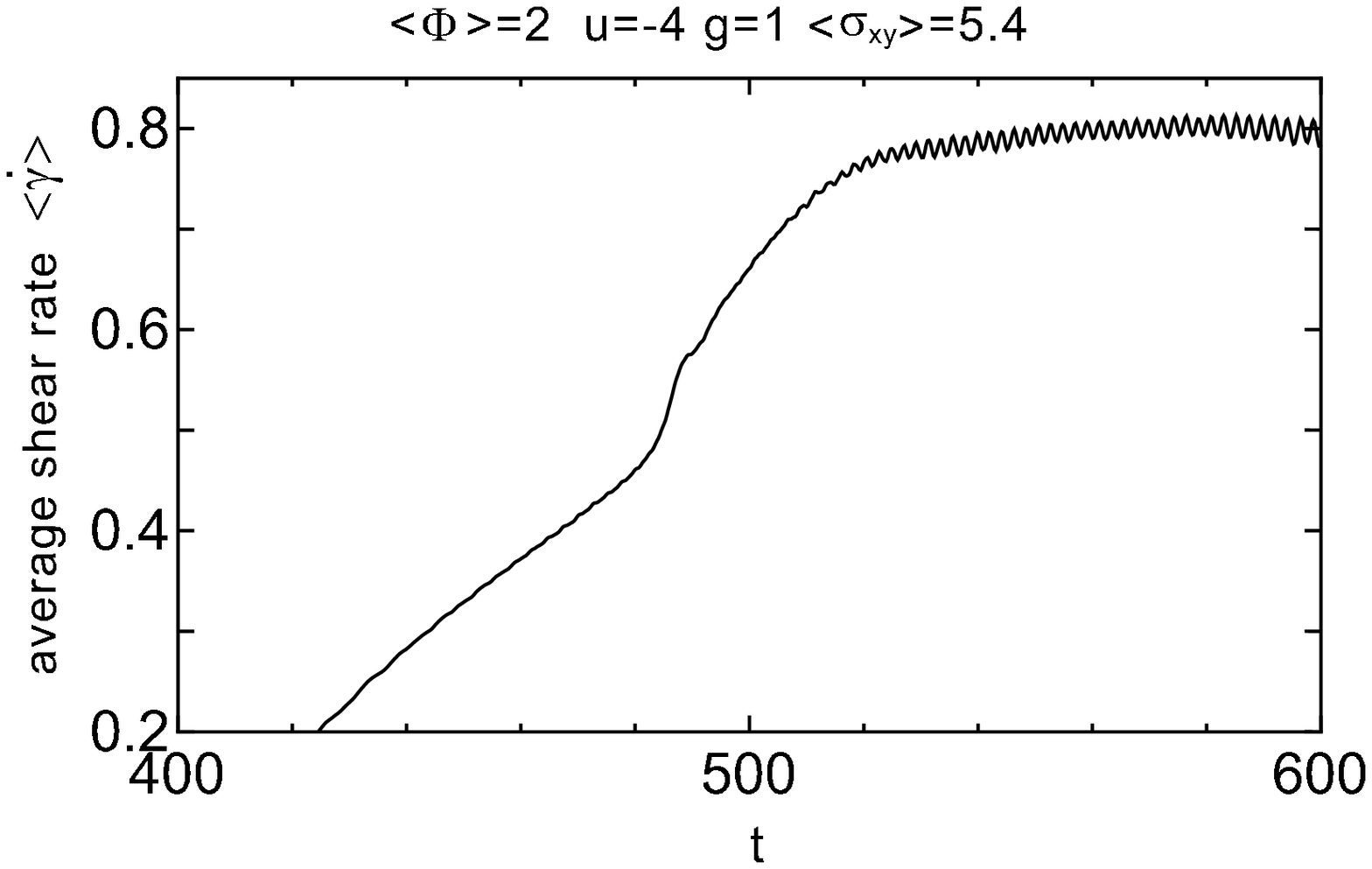}
  \end{center}
  \caption{\small{
      Time evolution of the average shear rate $\av{\gdot}(t)$ 
      for $\av{\sigma_{xy}}=5.4$ and $\av{\Phi}=2.0$.  
  }}
\end{figure}

\begin{figure}[h]
  \begin{center}
    \includegraphics[width=110mm]{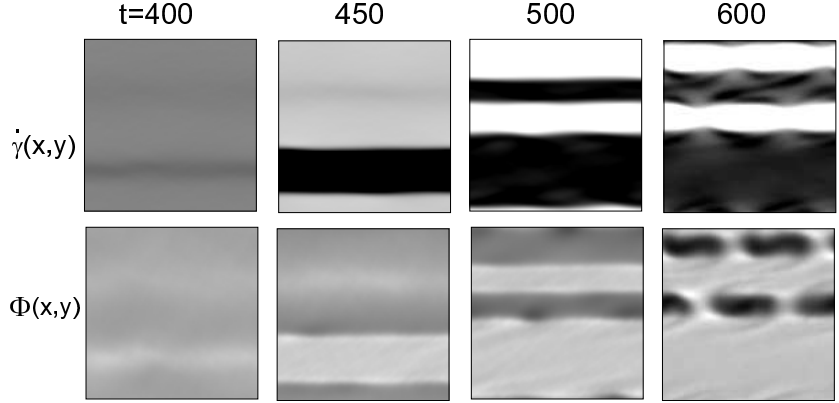}
  \end{center}
  \caption{\small{
      Snapshots of  $\Phi(\br,t)$ and $\gdot(\br,t)=\nb_y v_x$ 
      at $t=400$, 450, 500, 600. 
      The dark regions take larger values than the bright regions.  
  }}
\end{figure}
Because we set $u=-4$, 
 the system is in a one-phase  
state before application of shear. 
Figure 7 presents the time evolution of the average shear rate 
 $\av{\gdot}(t)= \av{\p v_x/\p y}$ in the initial stage of band 
formation after application of shear flow at $t=0$.   
In Fig.8 we show snapshots of 
$\Phi({\bi r},t)$ and $\gdot({\bi r},t)=
\nb_y v_x$ at various times in this stage. 
We can see that  the homogeneous state becomes unstable 
at $t\sim400$  
and  the shear bands subsequently begin to grow 
in the time region  $t=400\sim500$.   
The pseudo one-dimensional band structures 
 are soon deformed 
in the flow direction, 
resulting in the irregular 
oscillatory behavior of $\av{\gdot}(t)$.

We next explain the long-time behavior. 
Figure 9 shows the chaotic temporal fluctuations  
of the space average  $\av{\gdot}(t)= \av{\p v_x/\p y}$  
for $\av{\sigma_{xy}}=5.4$ and $\av{\Phi}=2.0$.  
We notice that $\av{\gdot}(t)$ frequently exhibits 
large amplitude fluctuations, 
which arises from large-scale disturbances  
of the interfaces  between different shear bands. 
Figure 10 illustrates $\Phi({\bi r},t)$ and $\gdot({\bi r},t)
=\nb_y v_x$ 
at $t=9000$ (A), $t=14000$ (B), and $t=15250$ (C) in Fig.9. 
The bird views of (A) and (C) are given also.   
 We find that the system jumps between the two states, 
(A) and (B), alternatively. In these states  
 the thickness of the low-shear band (white) is considerably 
different. 
Furthermore, at $t\sim15000$ in (C) 
the interface between two bands becomes globally unstable, 
resulting in abrupt expansion of the low-shear band.  
 We can see that the velocity 
and the composition are strongly  coupled. Thus 
these chaotic  spatio-temporal behavior arise from 
the stress-diffusion coupling  in  our model. 

\begin{figure}[htb]
  \begin{center}
    \includegraphics[width=100mm]{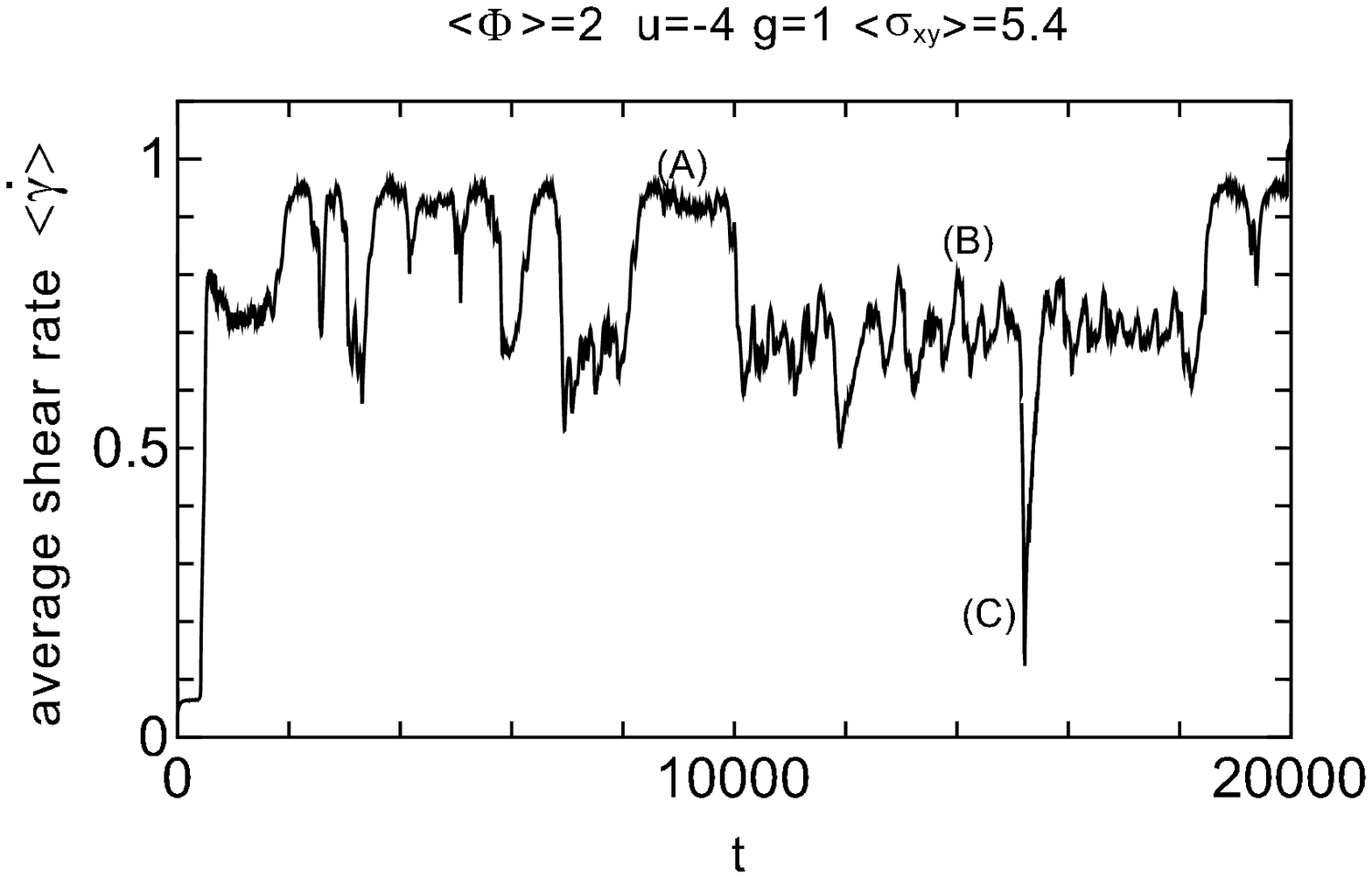}
    \label{xx}
  \end{center}
  \caption{\small{
      Time evolution of the average shear rate $\av{\gdot}(t)$ 
    for $\av{\sigma_{xy}}=5.4$ and $\av{\Phi}=2.0$.  
}}
\end{figure}
\begin{figure}[htb]
  \begin{center}
    \includegraphics[width=110mm]{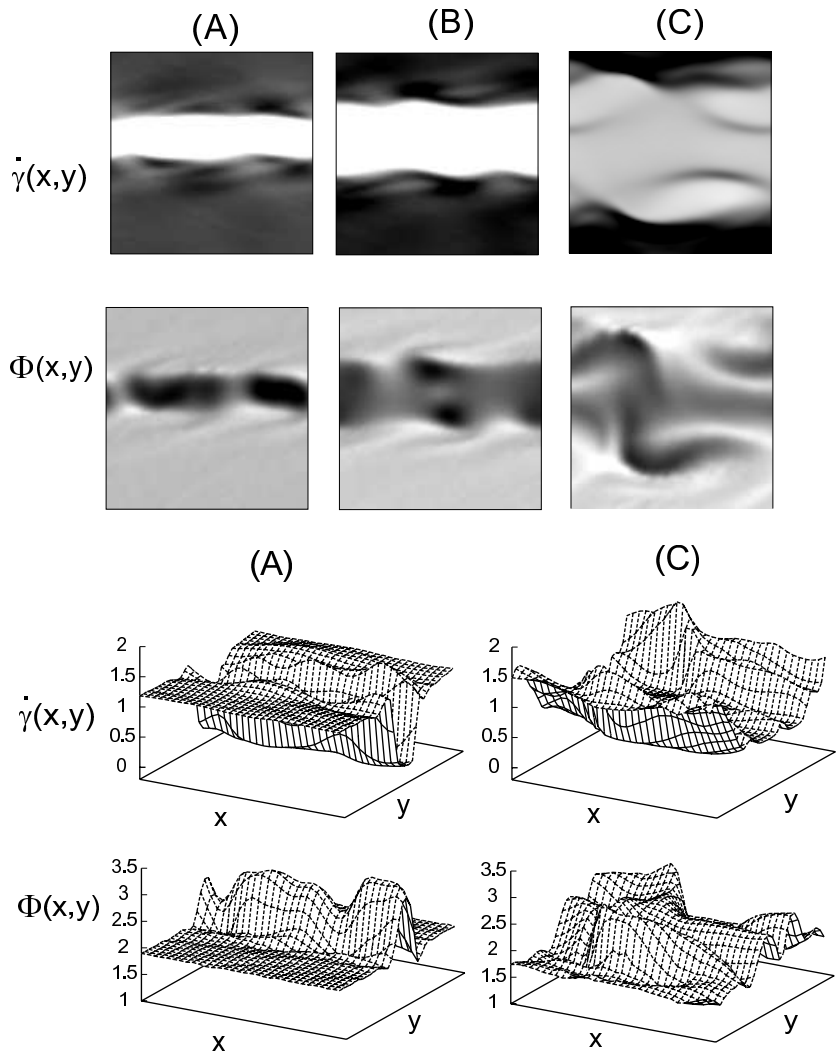}
    \label{xx}
  \end{center}
  \caption{\small{
    Snapshots and bird views 
    of  $\Phi(\br,t)$ and $\gdot(\br,t)=\nb_y v_x$ 
    at $t=9000$ (A), $t=14000$ (B), and $t=15250$ (C) in Fig.9. 
    The dark regions take larger values than the bright 
    regions  in the upper snapshots.  
    }}
\end{figure}

\section{Summary}

We have investigated complex spatio-temporal structures 
in sheared polymer systems by  solving 
a time-dependent Ginzburg-Landau model.   
Though performed in two dimensions, 
we have revealed a number of 
 unique features of sheared polymer systems. 
We  summarize our main results.\\ 
(i) In polymer solutions above the coexistence curve, 
Fig.4 shows  two dynamic regimes.  
In the linear regime  $\gdot<\gdot_{\rm nl}$ the shear-induced fluctuations 
can be maintained only in the presence of the thermal noise, 
while in the nonlinear regime $\gdot>\gdot_{\rm nl}$ 
chaotic dynamical steady states are realized due to the nonlinear 
interactions among the fluctuations. 
The previous linear calculations 
\cite{Helfand,OnukiPRL,Milner,Doi-Onuki} 
are meaningful only in the linear regime, 
and our results in Figs.1-5 are those in the nonlinear regime. 
The stress overshoot in Fig.1 
occurs due to appearance of the shear-induced 
composition fluctuations  in Fig.2. 
In the dynamical steady states of the nonlinear regime 
the heterogeneous deviation $\delta\phi$ is 
even on the order of the average $\av{\phi}$ 
varying on the scales of 
the characteristic size $\sim R_{\rm h}$ 
as  in the top figures of Fig.2. 
We have related  $R_{\rm h}$  
 to the average 
shear stress as in Eq.(12) 
 in the dynamical steady states.\\ 
(ii) We have also presented our preliminary 
results of shear-band structures 
in wormlike micellar solutions under the condition of fixed stress. 
As shown in Figs.9 and 10, 
the average shear rate exhibits large temporal fluctuations.  
However, we have not yet understood the mechanism 
why the shear bands exhibit irregular  behavior in 
shear flow.

We will shortly report 
 simulations of wormlike micellar solutions 
both at  fixed 
shear rate and at fixed shear stress. 
 We are not aware of any previous  
simulations at fixed  shear stress.    
 As other problems not treated here, we 
 mention  rheological turbulence   
 at low Reynolds numbers    in  dilute polymer solutions 
\cite{Steinberg} 
and  defect turbulence  in nematic 
liquid crystalline polymers  in shear flow 
 \cite{Onukibook,Mar,LD}. Both shear banding and tumbling 
have been observed in suspensions of 
rods \cite{Dhont}.  
Liquid crystalline  molecules can 
 tumble  in shear flow,  
constituting a coupled oscillator system   \cite{Kuramoto},  
 and loss of their phase 
coherency  leads  to 
proliferation of disclinations.  
Thus a number of nonequilibrium  problems 
   remain largely unexplored in 
complex fluids, which could  also be interesting subjects 
in  nonlinear science and in fluid mechanics.

{\bf Acknowledgments}\\

It is our great pleasure to acknowledge 
the lasting contribution of  Professor Yoshiki Kuramoto 
in nonlinear science   \cite{Kuramoto}.  
This work was supported by 
Grants in Aid for Scientific 
Research 
and for the 21st Century COE project 
(Center for Diversity and Universality in Physics)
 from the Ministry of Education, 
Culture, Sports, Science and Technology of Japan.

\end{document}